\documentclass[fleqn,10pt]{wlscirep}
\usepackage[utf8]{inputenc}
\usepackage[T1]{fontenc}

\usepackage[normalem]{ulem} 

\title{Trigonal Quasicrystalline States in $30^\circ$ Rotated Double Moir\'{e} Superlattices}

\author[1,2]{J. A. Crosse}
\author[1,2,3,*]{Pilkyung Moon}
\affil[1]{New York University Shanghai, Arts and Sciences, Shanghai, 200122, China}
\affil[2]{NYU-ECNU Institute of Physics at NYU Shanghai, Shanghai, 200062, China}
\affil[3]{New York University, Department of Physics, New York, 10003, USA}

\affil[*]{pilkyung.moon@nyu.edu}

\begin{abstract}
We study the lattice configuration and electronic structure
of a double moir\'{e} superlattice,
which is composed of a graphene layer encapsulated by
two other layers
in a way such that the two hexagonal moir\'{e} patterns are
arranged in a dodecagonal quasicrystalline configuration.
We show that there are between 0 and 4 such configurations
depending on the lattice mismatch between
graphene and the encapsulating layer.
We then reveal the resonant interaction,
which is distinct from the conventional 2-, 3-, 4-wave mixing
of moir\'{e} superlattices,
that brings together and hybridizes
twelve degenerate Bloch states of monolayer graphene.
These states do not fully satisfy the dodecagonal quasicrystalline rotational symmetry
due to the symmetry of the wave vectors involved.
Instead, their wave functions exhibit trigonal quasicrystalline order,
which lacks inversion symmetry,
at the energies much closer to the charge neutrality point of graphene.

\end{abstract}

\def\Vec#1{{\bf #1}}

\begin{document}

\flushbottom
\maketitle
%
%
\thispagestyle{empty}

\section*{Introduction}

When two or more two-dimensional atomic layers which do not share a common periodicity
are overlaid, an
additional periodicity in the form of
moir\'{e} interference pattern emerges \cite{Berger1191}.
The electronic structures of such systems - for example
twisted bilayer graphenes \cite{PhysRevLett.99.256802,PhysRevLett.101.056803,Bistritzer12233,moon2012energy,Moon2013},
graphene on hexagonal boron nitride (hBN) \cite{dean2010boron,Hunt1427,ponomarenko2013cloning,PhysRevB.86.115415,PhysRevB.86.081405,PhysRevB.87.245408,PhysRevB.90.155406},
and twisted bilayer transition metal dichalcogenides \cite{hsu2014second,fang2014strong}
with small twist angles $\theta \approx 1^\circ$ -
have been investigated extensively.
These materials have very long moir'{e} superlattice vectors
$\Vec{L}_i^\mathrm{M}$ ($i=1,2)$,
and, hence, exhibit many exotic properties
such as the Fermi velocity renormalization \cite{PhysRevLett.99.256802,PhysRevLett.101.056803},
mini Dirac points formation \cite{PhysRevB.86.081405,PhysRevB.90.155406},
Hofstadter's butterfly \cite{PhysRevB.84.035440,moon2012energy,dean2010boron,Hunt1427,ponomarenko2013cloning},
the emergence of superconductivity \cite{cao2018unconventional}, correlated phases \cite{cao2018correlated},
and orbital magnetic moment \cite{moriya2020emergence}.

\begin{figure}[ht]
\centering
\includegraphics[width=0.9\linewidth]{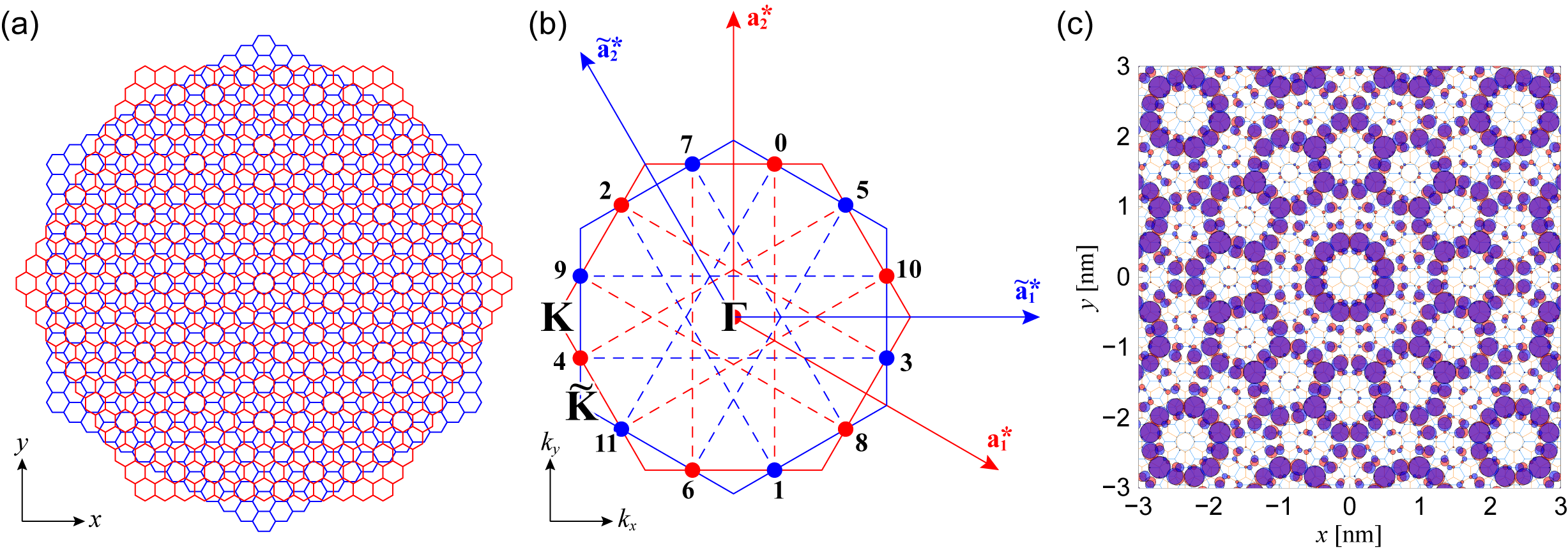}
\caption{
        (a) Lattice structures of quasicrystalline twisted bilayer graphene
        \cite{Ahn2018,PhysRevB.99.165430,PhysRevB.103.045408}.
		The red and blue hexagons represent the unit cells of each layer.
		(b) The wave vectors of the twelve monolayer states $\Vec{C}_n$ ($n=0,1,\cdots,11$)
		which hybridize to quasicrystalline resonant states.
		The red and blue hexagons, and the red and blue arrows [$\Vec{a}_i^*$ and $\tilde{\Vec{a}}_i^*$ ($i=1,2$)]
		represent the first Brillouin zones and the reciprocal lattice vectors of each layer.
		Due to the symmetry, these twelve states are all degenerate in energy,
		and the dashed lines show that these twelve states interact by the reciprocal lattice vectors
		of the two layers.
		Note that these twelve states are centered around the $\Gamma$ point.
		(c) Local density of states of the quasicrystalline resonant states.
		The area of the circle is proportional to the squared wave amplitude, and red and blue
circles represent the states in the upper and the lower layers, respectively.
}
\label{Figure_01}
\end{figure}

A special case occurs
when two hexagonal lattices are overlapped
at $\theta=30^\circ$ [Fig.~\ref{Figure_01}(a)]. In this instance
the atomic arrangement is mapped on to a quasicrystalline lattice,
which is ordered but not periodic,
with a 12-fold rotational symmetry \cite{Yao6928,Ahn2018,PhysRevB.99.165430,PhysRevB.100.081405,Yan_2019,yu2019dodecagonal,deng2020interlayer,PhysRevB.102.045113,PhysRevB.103.045408}.
Owing to the momentum mismatch \cite{Ahn2018}, quasicrystalline twisted bilayer graphene
exhibits the electronic structures of almost
decoupled bilayer graphene
at most energy ranges. Nevertheless,
it also hosts unique electronic states
which satisfy the 12-fold rotational symmetry
\cite{PhysRevB.99.165430,PhysRevB.102.045113,PhysRevB.103.045408}.
Such quasicrystalline states arise from
the resonant interaction 
between the states
at specific wave vectors
via the rotational symmetry of the quasicrystal
as well as the translational symmetry of
the constituent atomic layers
\cite{PhysRevB.99.165430,PhysRevB.103.045408}.
The red and blue hexagons in Fig.~\ref{Figure_01}(b)
show the first Brillouin zones of the two lattices. The numbered points and dashed lines
show the wave vectors of the constituent monolayer states
and the interlayer interaction
which form the quasicrystalline resonant states.
Such quasicrystalline states
exhibit a wave amplitude distributed selectively
on a limited number of sites in a characteristic 12-fold
rotationally symmetric pattern [Fig.~\ref{Figure_01}(c)].
These states, however,
appear at the energies (about $\pm1.7~\mathrm{eV})$ -
far from the charge neutrality point of graphene.
Similar quasicrystalline resonant states also
arise in any bilayer stacked in a quasicrystalline configuration
if all the dominant interlayer interactions occur between the atomic orbitals
that have the same magnetic quantum number \cite{PhysRevB.103.045408}.
Thus, even transition metal dichalcogenides or
square lattices can show the quasicrystalline states.

\begin{figure}[ht]
\centering
\includegraphics[width=0.9\linewidth]{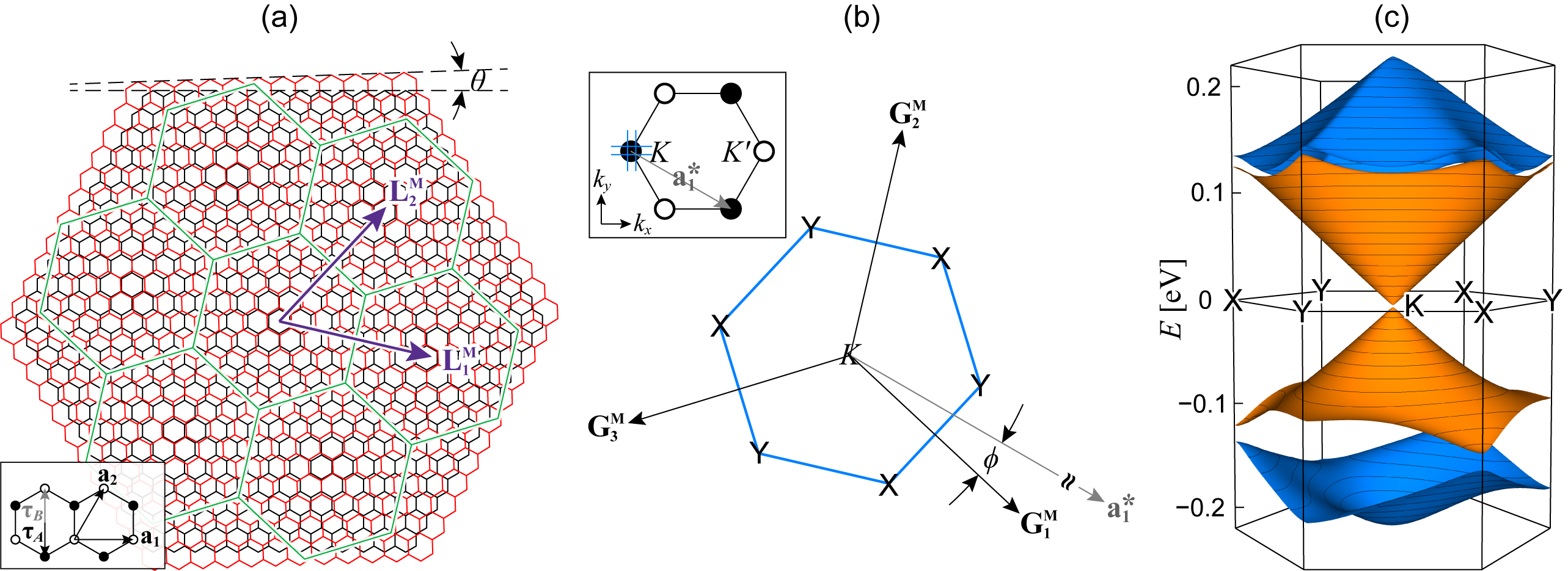}
\caption{
(a) Lattice structure of graphene on hBN \cite{Hunt1427,PhysRevB.90.155406}.
The black and red hexagons represent the unit cells of graphene and hBN, respectively,
and $\theta$ shows the relative orientation.
The green hexagons and the vectors $\Vec{L}_i^\mathrm{M}$ ($i=1,2$)
represent the unit cells and superlattice vectors of the moir\'{e} superlattice, respectively.
Here, the lattice constant of hBN is drawn 15\%
larger than that of graphene to enhance the visibility of the moir\'{e} pattern
(actual difference is about 1.79\%).
Inset shows the lattice configuration of graphene;
the black and white circles represent the $A$ and $B$ sublattices,
$\Vec{a}_i$ and $\boldsymbol{\tau}_X$ show the primitive lattice vectors
and the coordinates of the sublattices, respectively.
(b) Superlattice Brillouin zone (blue hexagon) near the Dirac point
(the region surrounded by blue lines in the inset)
and the reciprocal lattice vectors $\Vec{G}_i^\mathrm{M}$
of graphene on hBN.
$X$ and $Y$ show the Brillouin zone corners where mini Dirac point appear,
and $\phi$ shows the relative orientation of $\Vec{G}_1^\mathrm{M}$ to
the reciprocal lattice vector $\Vec{a}_1^*$ of pristine graphene.
Inset shows the first Brillouin zone of graphene,
where the black and white circles represent the three equivalent Dirac points, $K$ and $K'$, respectively.
(c) The band dispersion of the first two bands in the conduction and valence bands
of graphene on hBN with $\theta=0^\circ$,
which show the band opening at the primary and the mini Dirac points.
}
\label{Figure_02}
\end{figure}

Recently, rapid progress has been made
in stacking more than two incommensurate atomic layers and a number of studies have 
investigated the effects of multiple moir\'{e} superlattice potentials
on the electronic structure.
The most notable example among them is a double moir\'{e} system,
which is composed of a graphene layer encapsulated by
hBN layers (BN/G/BN) \cite{wang2019new,doi:10.1021/acs.nanolett.9b04058}.
The lattice mismatch between graphene and hBN
results in a hexagonal moir\'{e} superlattice potential
with a superlattice period $\Vec{L}_i^\mathrm{M}$ ($i=1,2$) that
can be as long as $14~\mathrm{nm}$
[Fig.~\ref{Figure_02}(a)].
Such a long period 
[which results in short superlattice reciprocal lattice vectors,
Fig.~\ref{Figure_02}(b)]
carves the graphene electronic structures
into superlattice bands with an energy scale much smaller than that of pristine graphene
[Fig.~\ref{Figure_02}(c)] \cite{PhysRevB.90.155406}.
Recently, Leconte and Jung show that BN/G/BN at specific configurations
can host two hexagonal moir\'{e} patterns
overlaid at a twist angle of $30^\circ$,
and claimed that the system hosts quasicrystalline electronic structures
\cite{Leconte_2020}.
However, 
the interaction mechanism responsible for such unique electronic states in BN/G/BN,
as well as the actual electronic band structures,
and whether the wave functions actually satisfy the symmetry of the quasicrystal
have not yet been investigated.

Here,
we investigate the conditions
where the two hexagonal \textit{moir\'{e} patterns} in double moir\'{e} superlattice
are arranged in a dodecagonal (12-fold) quasicrystalline configuration.
Then we reveal the resonant interactions
that bring together and hybridize twelve degenerate Bloch states of monolayer graphene
and show that such interactions
reconstruct the band dispersion of pristine graphene at these wave vectors.
Compared to the resonant states of quasicrystalline twisted bilayer graphene
where the actual \textit{atomic lattices} are arranged in a dodecagonal configuration
\cite{PhysRevB.99.165430,PhysRevB.102.045113,PhysRevB.103.045408},
the resonant states of BN/G/BN appear at the energies
much closer to the charge neutrality point of graphene.
However, their wave functions show the quasicrystalline order with a 3-fold rotational symmetry
rather than fully satisfying the 12-fold rotational symmetry of
the double moir\'{e} pattern.

\section*{Methods}

\subsection*{Hexagonal moir\'{e} superlattices stacked at $30^\circ$}

\begin{figure}[ht]
\centering
\includegraphics[width=0.8\linewidth]{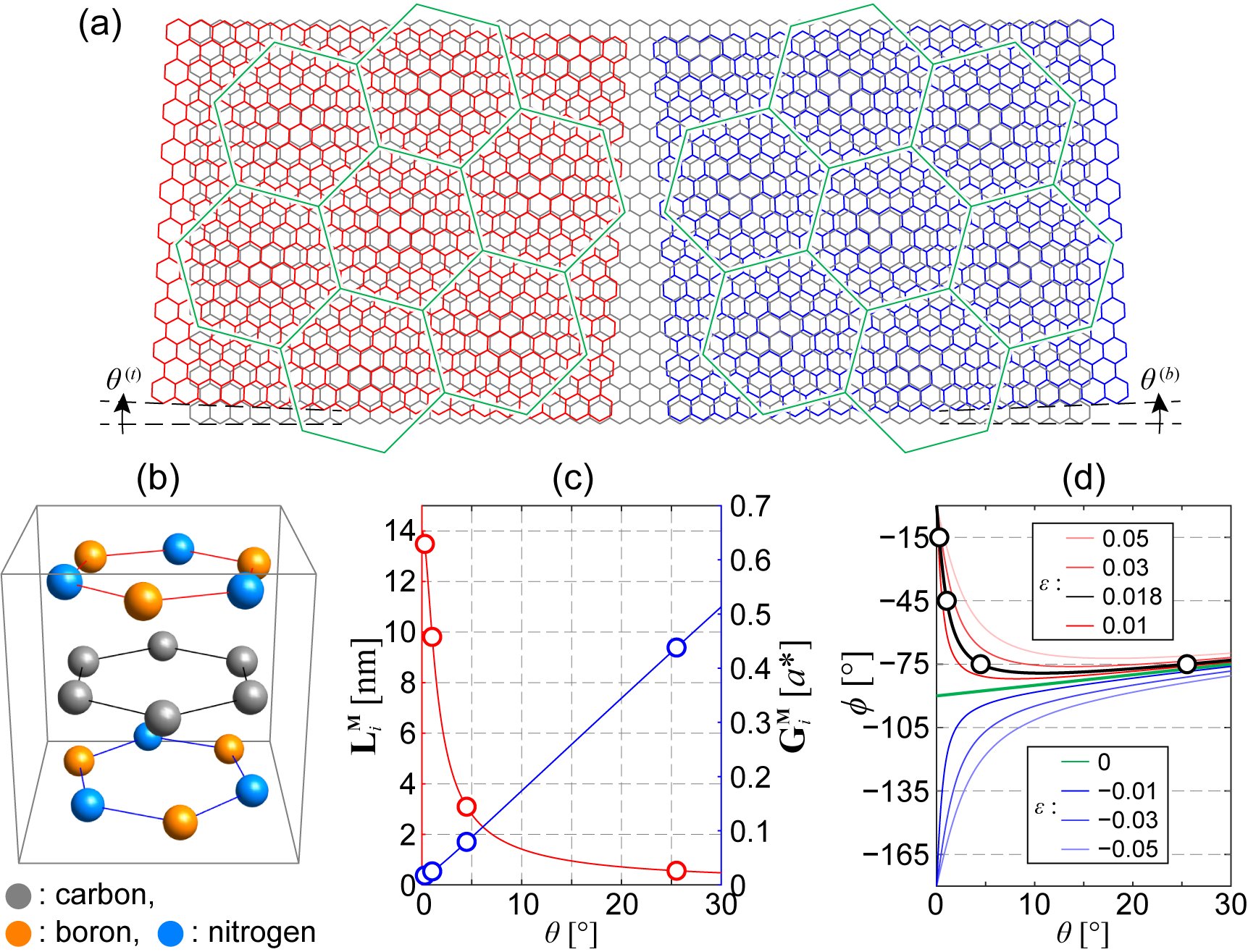}
\caption{
Lattice structure of graphene (gray hexagons)
encapsulated by the top (red) and bottom (blue) hBN layers
with twist angles of $\theta^{(t)}$ and $\theta^{(b)}$, respectively.
Green hexagons show the unit cells of the moir\'{e} superlattices
which are formed between the graphene and each of the hBN layer.
We draw only the top (bottom) hBN layer at the left (right) side
to enhance the visibility of the pattern.
The two hexagonal superlattice unit cells are
arranged at a relative angle of $30^\circ$,
and form a dodecagonal quasicrystalline pattern when overlaid.
(b) Atomic configuration of the three layers
which shows the $D_3$ point group symmetry.
(c) The lengths of the moir\'{e} lattice vectors (red line) and reciprocal lattice vectors (blue line)
plotted against $\theta$.
The circles correspond to the values
at the configuration shown in Fig.~\ref{Figure_03}(a).
(d) The angle $\phi$ between the primitive vectors
(both the real-space and the reciprocal lattice vectors)
of graphene and moir\'{e} superlattice plotted against
the twist angle $\theta$ between the two lattices.
The red, green and blue lines show the plot for the systems with
$\varepsilon>0$, $\varepsilon=0$ and $\varepsilon<0$, respectively.
The black line corresponds to that between graphene and hBN ($\varepsilon=0.0179$).
Black circles show the configuration
that can form the quasicrystalline arrangement of the moir\'{e} patterns
if the top and bottom hBN are rotated by $-\theta$ and $\theta$ from graphene, respectively.
}
\label{Figure_03}
\end{figure}

We consider a trilayer system composed of graphene sandwiched by hBN.
Both graphene and hBN are two-dimensional honeycomb lattice
whose unit cell comprises of two ($A$ and $B$) sublattices.
Graphene has carbon atoms in both sublattices,
while hBN has nitrogen atom on $A$ site and boron atom on $B$ site.
The lattice constant of hBN, $\tilde{a} \approx 0.2504~\mathrm{nm}$\cite{PhysRevB.68.104102},
is slightly larger than that of graphene, $a \approx 0.246~\mathrm{nm}$,
and we use a constant interlayer distance of $d=0.322\,\mathrm{nm}$
between the adjacent two layers\cite{PhysRevB.76.073103}.
Here, we do not consider
the lattice relaxation between graphene and hBN \cite{woods2014commensurate,jung2015origin},
since the effects of such relaxation on the electronic structures
is an order of a few meV.
Nevertheless, our effective theory that respects the lattice symmetry
is able to properly describe both
the gap at the primary Dirac point
and the asymmetric gap opening at the two inequivalent mini Dirac points,
\cite{PhysRevB.90.155406}
as well as the orbital magnetism of the structure \cite{moriya2020emergence}.

We define the atomic structure of the double moir\'{e} superlattices
by starting from a nonrotated arrangement,
where the hexagon center of the three layers share the same
in-plane position $(x,y)=(0,0)$,
and the $A$-$B$ bonds are parallel to each other.
We choose $\Vec{a}_1 = a(1,0)$ and $\Vec{a}_2=a(1/2,\sqrt{3}/2)$
($a=0.246\,\mathrm{nm}$)
as the primitive lattice vectors of graphene,
and $\boldsymbol{\tau}_A=-\boldsymbol{\tau}_1$ and $\boldsymbol{\tau}_B=\boldsymbol{\tau}_1$
[$\boldsymbol{\tau}_1 = -(1/3)(\Vec{a}_1-2\Vec{a}_2)$]
as the coordinates of the $A$ and $B$ sublattices in the unit cell.
The primitive lattice vectors of
the top ($l=t$) and the bottom ($l=b$) hBN layers become
$\tilde{\Vec{a}}_i^{(l)} = M \Vec{a}_i$ ($i=1,2$),
where $M= (1+\varepsilon) \mathbb{I}$ represents the isotropic expansion
by the factor $1+\varepsilon=\tilde{a}/a\approx 1.0179$,
and $\boldsymbol{\tau}_\mathrm{N}^{(l)}=-\boldsymbol{\tau}_1^{(l)}\pm d \Vec{e}_z$
and $\boldsymbol{\tau}_\mathrm{B}^{(l)}=\boldsymbol{\tau}_1^{(l)} \pm d\Vec{e}_z$
[$\boldsymbol{\tau}_1^{(l)} = -(1/3)(\tilde{\Vec{a}}_1^{(l)}-2\tilde{\Vec{a}}_2^{(l)})$],
where the upper and lower signs are for the top and bottom layers, respectively,
represent the coordinates of the nitrogen and boron atoms in the unit cell
\cite{comment:tau}.
We define the reciprocal lattice vectors $\Vec{a}_i^*$ and $\tilde{\Vec{a}}_i^*$
for graphene and hBN, respectively, so as to satisfy
$\Vec{a}_i \cdot \Vec{a}_j^* = \tilde{\Vec{a}}_i \cdot \tilde{\Vec{a}}_j^* = 2\pi \delta_{ij}$.
We then rotate the top and bottom hBN layers with respect to graphene by
arbitrary angles $\theta^{(t)}$ and $\theta^{(b)}$ around the origin, respectively.
From now on, we use "BN/G/BN" for this configuration only.
Due to the symmetry of the lattice,
$0\le \theta^{(l)} \le 30^\circ$ ($l\in t,b$) spans
all the independent configurations.

Figure \ref{Figure_03}(a) shows the moir\'{e} interference patterns
which arise from the lattice mismatch between the top hBN and graphene (left side),
and also that from the bottom hBN and graphene (right side), respectively,
and Fig.~\ref{Figure_03}(b) shows the atomic configuration of the three layers.
The lattice vectors $\Vec{L}_i^{\mathrm{M},(l)}$ and
the reciprocal lattice period $\Vec{G}_i^{\mathrm{M},(l)}$
($i=1,2$) of each moir\'{e} superlattice are
\begin{align}
\Vec{L}_i^{\mathrm{M},(l)} &=c R(\phi^{(l)}) \Vec{a}_i, \nonumber\\
\Vec{G}_i^{\mathrm{M},(l)} &=c^{-1} R(\phi^{(l)}) \Vec{a}^*_i,
\label{eq:LiM_and_GiM}
\end{align}
respectively, where
$c=(1+\varepsilon)/\sqrt{\varepsilon^2+2(1+\varepsilon)(1-\cos\theta^{(l)})}$,
$\phi^{(l)} = \mathrm{arctan} [-\sin\theta^{(l)}/(1+\varepsilon-\cos\theta^{(l)})]$,
and $R(\phi)$ is a rotation by $\phi$
\cite{yankowitz2012emergence,PhysRevB.90.155406}.
We plot $|\Vec{L}_i^{\mathrm{M}}|$ and $|\Vec{G}_i^{\mathrm{M}}|$
against $\theta$ in Fig.~\ref{Figure_03}(c)
in red and blue lines, respectively.

\begin{figure}[ht]
\centering
\includegraphics[width=0.65\linewidth]{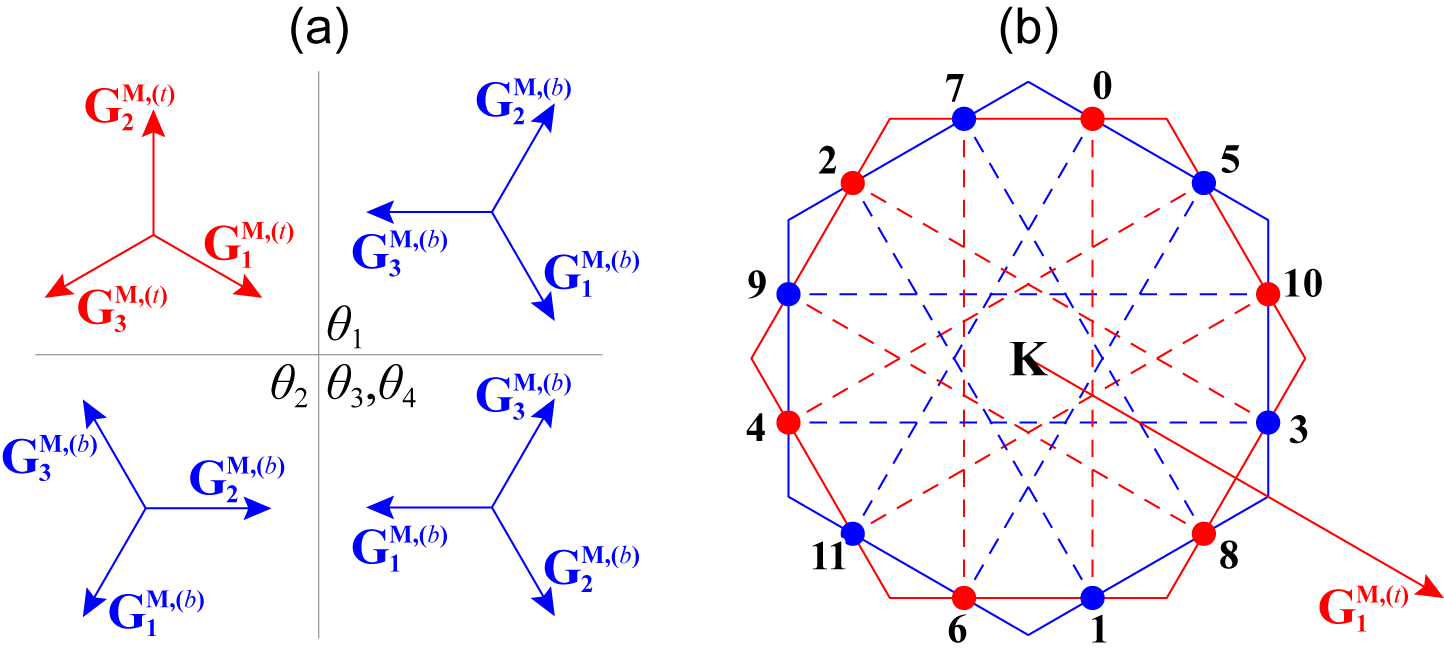}
\caption{
(a) Relative orientation of the moir\'{e} reciprocal lattice vectors
of the lower moir\'{e} superlattice ($\Vec{G}_i^{\mathrm{M},(b)}$, blue arrows)
for $\theta_i$ ($i=1,2,3,4$)
with respect to the direction of the vectors
of the upper moir\'{e} superlattice ($\Vec{G}_i^{\mathrm{M},(t)}$, red arrows).
(b) Relative direction of the wave vectors
involved in the resonant coupling [showing the number $n$ of $\Vec{C}_n$, Eq.~\eqref{eq:C_n}]
with respect to the direction of $\Vec{G}_i^{\mathrm{M},(t)}$.
In both figures,
note that the actual direction of $\Vec{G}_1^{\mathrm{M},(t)}$
is $\phi$ rotated from $\Vec{a}_1^*$ [Eq.~\eqref{eq:LiM_and_GiM}].
}
\label{Figure_04}
\end{figure}

Now, we will find the configuration
where the unit cells of the two hexagonal moir\'{e} superlattices
have the same size
and are overlaid with a relative twist angle of $30^\circ$.
In such a configuration, the overlaid two hexagonal superlattices
are mapped onto a 12-fold rotationally symmetric quasicrystalline lattice 
without any translational symmetry,
as first shown by Stampfli \cite{stampfli}. 
From Eq.~\eqref{eq:LiM_and_GiM},
the former and the latter conditions give $|\theta^{(t)}|=|\theta^{(b)}|$
and $\phi^{(t)}-\phi^{(b)} \equiv 30^\circ ~(\mathrm{mod}~60^\circ)$,
which can be simultaneously satisfied by
$\theta^{(t)}=-\theta^{(b)}$ and
$\phi^{(t)}=-\phi^{(b)}\equiv 15^\circ ~(\mathrm{mod}~30^\circ)$.
Figure \ref{Figure_03}(d) shows $\phi$ as a function of $\theta$
for various $\varepsilon$.
The red, green, blue lines correspond to
$\varepsilon>0$, $\varepsilon=0$, $\varepsilon<0$,
respectively, and the thick black line corresponds to hBN.
The two hexagonal moir\'{e} superlattices form
a dodecagonal quasicrystalline configuration
at $\theta$ where the line and the dashed horizontal lines cross.
If the lattice constant of the top and bottom layers is
the same as that of the middle, graphene layer, i.e., $\varepsilon=0$,
then the two hexagonal moir\'{e} patterns cannot have
a relative twist angle of $30^\circ$.
On the other hands, the systems with $\varepsilon<0$,
$0<\varepsilon < 0.0353$, and $\varepsilon \ge 0.0353$ have
three, four, and two $\theta$ which satisfy the conditions. 
When the top and bottom layers are hBN, i.e., $\varepsilon\approx 0.0179$,
the four angles are
$\theta_1=0.274^\circ$, $\theta_2=1.03^\circ$,
$\theta_3=4.48^\circ$, $\theta_4=25.5^\circ$,
and the corresponding $|\Vec{G}_i^\mathrm{M}|$ are
0.0182, 0.0251, 0.0795, 0.438 times the $|\Vec{a}_i^*|$.
Note that $\theta_4$ gives very long $|\Vec{G}_i^\mathrm{M}|$,
and accordingly very short $|\Vec{L}_i^\mathrm{M}|$,
which competes with the length scale of monolayer graphene.
By choosing $\theta^{(t)}=-\theta_i$ and $\theta^{(b)}=\theta_i$ ($i=1,2,3,4$),
we get $\phi^{(t)}=-\phi_i$ and $\phi^{(b)}=\phi_i$,
where $\phi_1=-15^\circ$, $\phi_2=-45^\circ$, and $\phi_3=\phi_4=-75^\circ$.
Then, the twelve moir\'{e} reciprocal lattice vectors
\begin{equation}
\{\pm \Vec{G}_i^{\mathrm{M},(l)} ~| ~i=1,2,3, ~l=t,b\},
\label{eq:12vectors}
\end{equation}
where $\Vec{G}_3^{\mathrm{M},(l)} = -\Vec{G}_1^{\mathrm{M},(l)}-\Vec{G}_2^{\mathrm{M},(l)}$,
are arranged in 12-fold rotational symmetry [Fig.~\ref{Figure_04}(a)],
just like the reciprocal lattice vectors in
quasicrystalline twisted bilayer graphene
that give rise to the resonant states [Fig.~\ref{Figure_01}(b)]\cite{PhysRevB.99.165430}.

It should be noted that, however, 
although the overlap of the two moir\'{e} interference patterns
are mapped onto a quasicrystalline tiling with 12-fold rotational symmetry,
the actual lattice structure belongs to the symmetry group $D_3$;
it is invariant under $C_3$ rotation about the axis perpendicular to the $xy$-plane and under three $C_2$ rotation about the axes in the plane,
but lacks inversion symmetry.
If we replace the top and bottom hBN layers
by a material having the same types of atoms in both sublattices,
then the lattice has the symmetry group $D_6$
which is still lower than the 12-fold rotational symmetry.

\subsection*{Hamiltonian of double moir\'{e} superlattices}

The total tight-binding Hamiltonian of the double moir\'{e} superlattice
is expressed as
\begin{equation}
    H = H_\mathrm{G} + H_{\mathrm{hBN}}^{(t)} + H_\mathrm{hBN}^{(b)} + U^{(t)} + U^{(l)},
\end{equation}
where $H_\mathrm{G}$ and
$H_\mathrm{hBN}^{(t)}$ ($H_\mathrm{hBN}^{(b)}$) represent
the Hamiltonian for the intrinsic monolayer graphene and
the top (bottom) hBN, respectively,
$U^{(t)}$ ($U^{(b)}$)
is for the interlayer coupling between
the graphene and the top (bottom) hBN.
However, since the hBN electronic bands are far from
the charge neutrality point of graphene,
we can project the total Hamiltonian
on to the Bloch bases of graphene $p_z$ orbitals at each sublattice,
\begin{equation}
	|\Vec{k},X\rangle = 
	\frac{1}{\sqrt{N}}\sum_{\Vec{R}_{X}} e^{i\Vec{k}\cdot\Vec{R}_{X}}
	|\Vec{R}_{X} \rangle
\end{equation}
where $|\Vec{R}_{X} \rangle$ is the atomic orbital at the site
$\Vec{R}_{X}=n_1 \Vec{a}_1 + n_2 \Vec{a}_2 + \boldsymbol{\tau}_X$
($n_i \in \mathbb{Z}$, $X=A,B$), 
$\Vec{k}$ is the two-dimensional Bloch wave vectors
and $N = S_\mathrm{tot}/S$ is the number of the graphene unit cells
with an area $S=(\sqrt{3}/2)a^2$
in the total system area $S_\mathrm{tot}$.
Then, the Hamiltonian near the Dirac point $K^\xi = -\xi(2\Vec{a}_1^*+\Vec{a}_2^*)/3$
of graphene,
where $\xi=\pm1$ for $K$ and $K'$, respectively,
is reduced to a $2\times 2$ form
\cite{PhysRevB.90.155406,comment:for_atomic_layers_other_than_hBN},
\begin{align}
 \tilde{H} &= 
H_\mathrm{G} + U^{(t)\dagger} (-H_{\rm hBN}^{(t)})^{-1} U^{(t)}+ U^{(b)\dagger} (-H_{\rm hBN}^{(b)})^{-1} U^{(b)}
\nonumber\\
&\equiv
H_{\rm G} +  V_{\rm hBN}^{(t)}+  V_{\rm hBN}^{(b)}.
\label{eq:reduced_H}
\end{align}
The intralayer matrix elements of graphene are given by
\begin{align}
H_\mathrm{G}&=
\begin{pmatrix}
h_{AA} & h_{AB} \\ h_{BA} & h_{BB}
\end{pmatrix},
\nonumber\\
 h_{X,X'}(\Vec{k})  &= \sum_{\Vec{L}}
-T(\Vec{L}+\boldsymbol{\tau}_{X'X}) e^{-i\Vec{k}\cdot(\Vec{L}+\boldsymbol{\tau}_{X'X})},
\label{eq_H0}
\end{align}
where $\Vec{L} = n_1 \Vec{a}_1 + n_2 \Vec{a}_2$,
$\boldsymbol{\tau}_{X'X} = \boldsymbol{\tau}_{X'}- \boldsymbol{\tau}_{X}$, and
\begin{align}
-T(\Vec{R}) &= 
V_{pp\pi}\left[1-\left(\frac{\Vec{R}\cdot\Vec{e}_z}{R}\right)^2\right]
+ V_{pp\sigma}\left(\frac{\Vec{R}\cdot\Vec{e}_z}{R}\right)^2,
\nonumber \\
V_{pp\pi} &=  V_{pp\pi}^0 e^{- (R-a/\sqrt{3})/r_0},
\quad V_{pp\sigma} =  V_{pp\sigma}^0  e^{- (R-d)/r_0},
\label{eq:transfer_integral}
\end{align}
is the transfer integral between two $p_z$ orbitals
at a relative vector $\Vec{R}$ \cite{slater_koster,moon2012energy},
$V_{pp\pi}^0 \approx -3.38~\mathrm{eV}$ \cite{comment_on_parameters},
$V_{pp\sigma}^0 \approx 0.48~\mathrm{eV}$,
and $r_0 \approx 0.0453\,\mathrm{nm}$ \cite{TramblydeLaissardiere2010,Moon2013}.
The effective potentials by hBN to graphene, $V_{\rm hBN}^{(l)}$, are explicitly written as
\cite{PhysRevB.90.155406,comment:tau}
\begin{equation}
 V_{\rm hBN}^{(l)} = W_0 + \{
 W_1^\xi e^{i\xi \Vec{G}_1^{\mathrm{M},(l)} \cdot \Vec{r}}+
 W_2^\xi e^{i\xi \Vec{G}_2^{\mathrm{M},(l)} \cdot \Vec{r}}+
 W_3^\xi e^{i\xi \Vec{G}_3^{\mathrm{M},(l)} \cdot \Vec{r}}+\mathrm{h.c.}\},
\label{eq:V_hBN}
\end{equation}
where we truncated much weaker terms $\mathcal{O}(u_0^4)$
which are associated with longer momentum difference. Here,
\begin{equation}
W_0 =
V_0
\begin{pmatrix}
1 & 0
\\
0 & 1
\end{pmatrix},\quad
W_1^\xi =
V_1 e^{i\xi\psi}
\begin{pmatrix}
1 & \omega^{-\xi}
\\
1 & \omega^{-\xi}
\end{pmatrix},\quad
W_2^\xi =
V_1 e^{i\xi\psi}
\begin{pmatrix}
1 & \omega^{\xi}
\\
\omega^{\xi} & \omega^{-\xi}
\end{pmatrix},\quad
W_3^\xi =
V_1 e^{i\xi\psi}
\begin{pmatrix}
1 & 1
\\
\omega^{-\xi} & \omega^{-\xi}
\end{pmatrix},
\label{eq:W0-W3}
\end{equation}
and
\begin{align}
  V_0 &= -3 u_0^2 
\left(
\frac{1}{V_{\rm N}}
+ \frac{1}{V_{\rm B}}
\right),
\nonumber\\
  V_1 e^{i\xi\psi} &= - u_0^2 
\left(
\frac{1}{V_{\rm N}}
+ \omega^\xi \frac{1}{V_{\rm B}}
\right),
\label{eq:V0_V1}
\end{align}
where $\omega = e^{2\pi i/3}$,
and $u_0 \approx -t(K^\xi) \approx 0.152~\mathrm{eV}$ is the in-plane Fourier transformation 
of the transfer integral between two $p_z$ orbitals
[Eq.~\eqref{eq:transfer_integral}]
\begin{eqnarray}
t(\Vec{q}) = 
\frac{1}{S} \int
T(\Vec{r}+ z_{\tilde{X}X}\Vec{e}_z) 
e^{-i \Vec{q}\cdot \Vec{r}} d\Vec{r}
\label{eq:t(q)}
\end{eqnarray}
at $\Vec{q}$ near the Dirac point \cite{slater_koster,moon2012energy}.
We will discuss more about $u_0$ later.
By using $V_\mathrm{C}=0$, $V_\mathrm{N}=-1.40~\mathrm{eV}$, and $V_\mathrm{B}=3.34~\mathrm{eV}$,
as the on-site potential of carbon, nitrogen, and boron atoms, respectively
\cite{PhysRevB.81.155433},
we get
$V_0 \approx 0.0289~\mathrm{eV}$,
$V_1 \approx 0.0210~\mathrm{eV}$,
and $\psi \approx -0.29$ (rad).
If we replace the top and bottom hBN layers
by a material having the same types of atoms in both sublattices,
i.e., if $V_\mathrm{N}=V_\mathrm{B}$,
then $\psi\equiv \pi/3  ~(\mathrm{mod}~\pi)$.
The symmetry of such a structure increases to $D_6$,
and the reduced Hamiltonian $\tilde{H}$ gains the inversion symmetry
\begin{equation}
    \tilde{H}^{(-\xi)}(\Vec{k},\Vec{r})=\sigma_x [\tilde{H}^{(\xi)}(-\Vec{k},-\Vec{r})] \sigma_x.
\end{equation}
It is straightforward to show that
the reduced Hamiltonian $\tilde{H}$ spans the subspace
\begin{equation}
\{|\Vec{k},X\rangle ~| ~ \Vec{k}=\Vec{k}_0+\sum\limits_{l=t,b} \sum\limits_{i=1,2,3}  m_i^{(l)} \Vec{G}_i^{\mathrm{M},(l)}, ~ m_i^{(l)}\in\mathbb{Z} \},
\label{eq:subspace}
\end{equation}
for any $\Vec{k}_0$ in the momentum space.
To investigate the electronic structures near $\Vec{k}_0$, for
any practical calculation, we only need a limited
number of bases around $\Vec{k}_0$ inside a certain cut-off circle $k_c$,
because the interaction with the states far from $\Vec{k}_0$
is very weak due to multiple scattering.
We can, then, obtain
the energy eigenvalues at all the wave vectors in Eq.~\eqref{eq:subspace}
by diagonalizing the Hamiltonian matrix within the finite bases,
and the quasiband dispersion of the system
by plotting the energy levels against $\Vec{k}_0$.
Here the wave number $\Vec{k}_0$ works like the crystal momentum
for the periodic system, and so it can be called the quasicrystal momentum.

\subsection*{Quasicrystalline resonant interaction by two moir\'{e} superlattice potentials}

In addition to the typical 2- and 3-wave interaction
by each moir\'{e} superlattice
[Fig.~\ref{Figure_02}(c)]\cite{PhysRevB.90.155406},
the 12-fold rotational symmetry of
the wave vectors
which couple the monolayer states [Eq.~\eqref{eq:12vectors}]
as well as the translational symmetries
of the two moir\'{e} superlattices
enables a unique interaction
between twelve degenerate monolayer states.
Such a resonant coupling occurs at the twelve symmetric points
\begin{equation}
    \Vec{C}_n =K^\xi + 2|\Vec{G}_i^{\mathrm{M},(l)}| \sin(\pi/12) (\cos \theta_n, \sin \theta_n),
    \label{eq:C_n}
\end{equation}
shown in Fig.~\ref{Figure_04}(b),
where $\theta_n = \frac{5\pi}{12} + \frac{7n\pi}{6} - \phi_i$ ($n=0,1,2,\cdots,11$).
While the twelve waves which constitute the resonant states
in quasicrystalline twisted bilayer graphene are centered around the $\Gamma$ point
\cite{PhysRevB.103.045408},
the waves involved in the resonant interaction in BN/G/BN
are centered around the Dirac point $K^\xi$.
These twelve states are degenerate
if we ignore the small trigonal warping in this low energy regime.
We see that the states at $\Vec{C}_i$
strongly interact with the states at $\Vec{C}_{i-1}$ and $\Vec{C}_{i+1}$
by the reciprocal lattice vectors of the top and bottom moir\'{e} superlattices.
The interaction to the states at any other $\Vec{k}$
can be safely neglected 
since the interaction strength is much less or
the two states are not degenerate in most cases.
Hence, these states form one-dimensional monatomic chain
with twelve sites and two pseudospins
whose interaction is
described by the moir\'{e} potential [Eq.~\eqref{eq:reduced_H}].


It should be noted that this is not the only resonant coupling in this system.
As shown in previous work on quasicrystalline twisted bilayer graphene
(e.g., Appendix A in Ref.~\cite{PhysRevB.103.045408}),
there are more sets of states, with different wave numbers, that
show the resonant interaction between the constituent monolayer states.
However, the set in Fig.~\ref{Figure_04}(b) is associated with
the strongest interaction $|t(\Vec{q})|$ and, hence, gives
the largest energy separation between the hybridized states.

\section*{Results and Discussion}




By using the Bloch bases ($|\Vec{k}^{(0)}\rangle$, $|\Vec{k}^{(1)}\rangle$, $~\cdots$, $|\Vec{k}^{(11)}\rangle$)
near the twelve wave vectors $\Vec{k}^{(n)}=\Vec{C}_n+\Vec{k}_0$,
where $|\Vec{k}^{(n)}\rangle$ is
$(|\Vec{k}^{(n)},A\rangle,\eta|\Vec{k}^{(n)},B\rangle)$
with $\eta=\omega^{\xi \times \mathrm{floor}(n/4)}$,
we can express the Hamiltonian of the resonant ring $H_{\mathrm{ring}}^\xi$
by a $24\times 24$ matrix
\setcounter{MaxMatrixCols}{20}  
\begin{align}
H_{\rm ring}^\xi(\Vec{k}_0) = 
\begin{pmatrix}
H_0^{(0)} & W_2^\xi &&&&&&&&&& Y_1^{\xi\dagger} \\
W_2^{\xi\dagger} & H_1^{(0)} & X_2^{\xi\dagger} \\
& X_2^\xi & H_2^{(0)} & W_1^{\xi\dagger} \\
&& W_1^\xi & H_3^{(0)} & Y_1^\xi \\
&&& Y_1^{\xi\dagger} & H_0^{(4)} & W_2^\xi \\
&&&& W_2^{\xi\dagger} & H_1^{(4)} & X_2^{\xi\dagger} \\
&&&&& X_2^{\xi} & H_2^{(4)} & W_1^{\xi\dagger} \\
&&&&&& W_1^{\xi} & H_3^{(4)} & Y_1^{\xi} \\
&&&&&&& Y_1^{\xi\dagger} & H_0^{(8)} & W_2^{\xi} \\
&&&&&&&& W_2^{\xi\dagger} & H_1^{(8)} & X_2^{\xi\dagger} \\
&&&&&&&&& X_2^{\xi} & H_2^{(8)} & W_1^{\xi\dagger} \\
Y_1^\dagger&&&&&&&&&& W_1^{\xi} & H_3^{(8)} \\
\end{pmatrix}.
\label{eq_dodeca_H_ring}
\end{align}
Here
\begin{align}
H_i^{(n)} &= 
\begin{pmatrix}
h_{AA}^{(n,i)} & h_{AB}^{(n,i)}  \\
h_{BA}^{(n,i)} & h_{BB}^{(n,i)}  \\
\end{pmatrix} + 2W_0, \nonumber\\
h_{X'X}^{(n,i)}(\Vec{k}_0) &= h_{X'X}[R(-7n\pi/6) \Vec{k}_0 + \Vec{C}_i],
\end{align}
and 

\begin{equation}
X_2^\xi = \left\{\begin{array}{l}
W_1^{\xi\dagger}\qquad \mathrm{for\;\theta_{1}}   \\
W_3^\xi \qquad \hspace{0.1cm}\mathrm{for\;\theta_{2}}  \\
W_2^{\xi\dagger} \qquad \mathrm{for\;\theta_{3}\;and\;\theta_{4}}  
\end{array}\right.\qquad\qquad
Y_1^\xi = \left\{\begin{array}{l} 
V_1 e^{-i\xi\psi} 
\begin{pmatrix}
1 & \omega^{-\xi}
\\
1 & \omega^{-\xi}
\end{pmatrix}\qquad\hspace{0.2cm}\mathrm{for\;\theta_{1}}   \\
V_1 e^{i\xi\psi} 
\begin{pmatrix}
1 & \omega^{-\xi}
\\
\omega^{\xi} & 1
\end{pmatrix}\qquad\hspace{0.15cm}\mathrm{for\;\theta_{2}}  \\
V_1 e^{-i\xi\psi} 
\begin{pmatrix}
1 & \omega^{\xi}
\\
\omega^\xi & \omega^{-\xi}
\end{pmatrix} \qquad \mathrm{for\;\theta_{3}\;and\;\theta_{4}}  
\end{array}\right.
\end{equation}
for $\theta_1$, $\theta_2$, and both $\theta_3$ and $\theta_4$, respectively.

\begin{figure}[ht]
\centering
\includegraphics[width=0.9\linewidth]{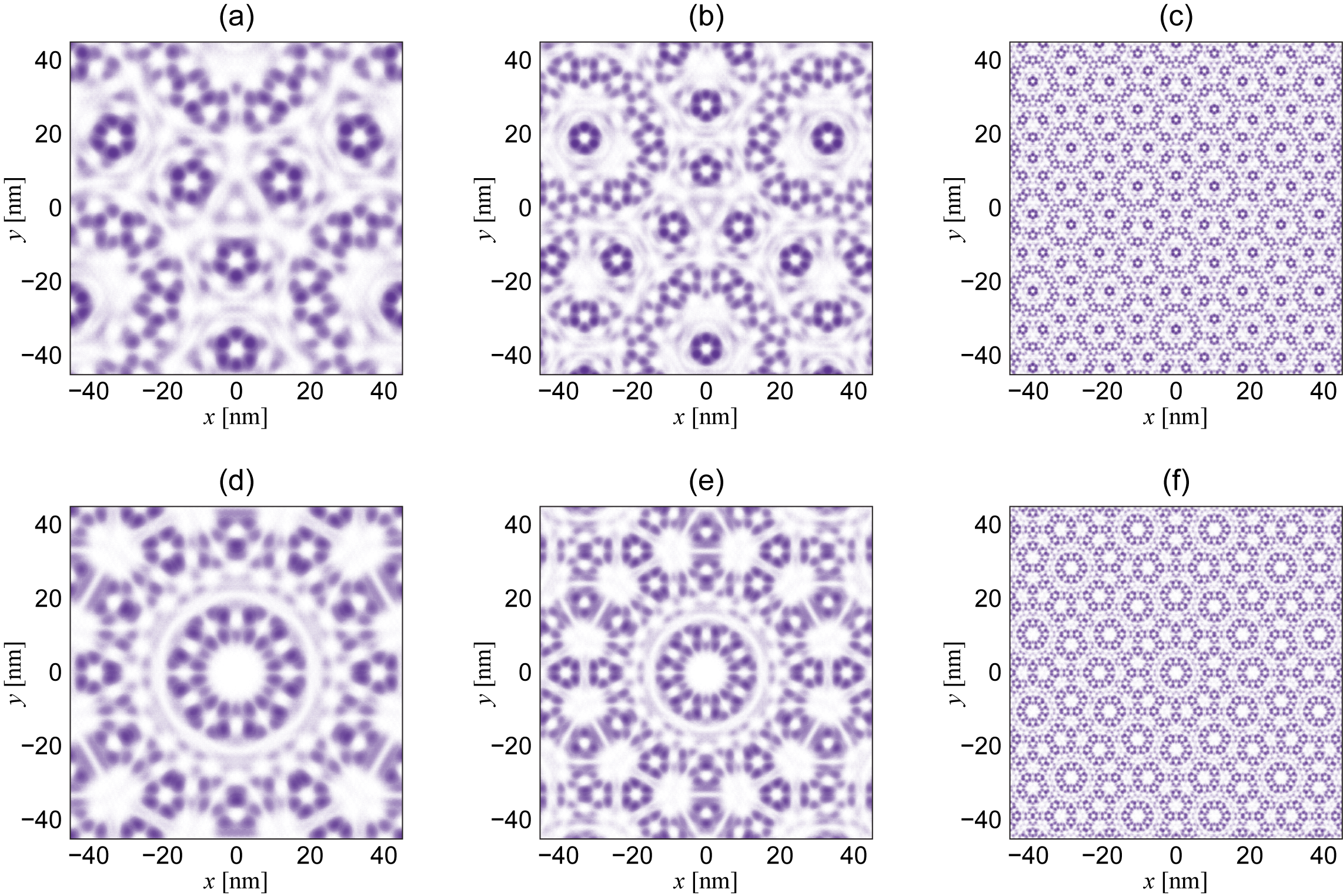}
\caption{
The electron probability distribution
of the wave functions at $\Vec{k}_0=\Vec{0}$
of the double moir\'{e} superlattice
(a)-(c) in the absence of the inversion symmetry,
i.e., $V_\mathrm{N}\ne V_\mathrm{B}$
and $\psi \neq \pi/3 ~(\mathrm{mod}~\pi)$, and
(d)-(f) in the presence of the inversion symmetry,
i.e., $V_\mathrm{N} = V_\mathrm{B}$
and $\psi \equiv \pi/3 ~(\mathrm{mod}~\pi)$.
(a) and (d) are for $\theta_1$,
(b) and (e) are for $\theta_2$, and
(c) and (f) are for $\theta_3$.
We show the third state ($m=\xi$) in the valence band,
since the first two states ($m=0,-\xi$) are degenerate.
}
\label{Figure_06}
\end{figure}

Obviously, Eq.~\eqref{eq_dodeca_H_ring} is
symmetric under rotation by \textit{four} span of the ring
(i.e., moving $\Vec{C}_n$ to $\Vec{C}_{n+4}$),
which actually corresponds to the
$[R(7\pi/6)]^4$ ($120^\circ$ rotation) of the entire system.
This means that 
the resonant states of the BN/G/BN have a 3-fold rotational symmetry,
which is much lower than the 12-fold rotational symmetry of the double moir\'{e} patterns
of the system.
This Hamiltonian cannot obtain a 6-fold rotational symmetry,
even if we replace the top and bottom hBN layers
by a material having the same types of atoms in both sublattices
(e.g., $V_\mathrm{N}=V_\mathrm{B}$),
since
(i) atomic structures lacks the 12-fold rotational symmetry and
(ii) the 12 wave vectors involved are centered at $K^\xi$ which has a 3-fold rotational symmetry
[Fig.~\ref{Figure_02}(b)].
Thus, the relevant terms cannot be gauged out by a similarity transformation.
There is, however, an exception in the systems with $V_\mathrm{N}=V_\mathrm{B}$;
the wave functions at $\textbf{k}_0=\textbf{0}$, and only at this $\textbf{k}_0$,
show a 6-fold rotational symmetry [Figs.~\ref{Figure_06}(d)-(f)].

The interlayer interaction strength $u_0$ [Eq.~\eqref{eq:V0_V1}]
deviates from $-t(K^\xi)$
as the distance between $\Vec{k}^{(n)}$ and $K^\xi$ increases.
This effect becomes obvious in the system with a longer $|\Vec{G}_i^{\mathrm{M},(l)}|$,
e.g., in BN/G/BN with $\theta_4$.
However, what is more important is that, 
the $u_0$ associated with 
the interaction between the neighboring Bloch states $|\Vec{k}^{(n)}\rangle$ ($n=0,1,\cdots,11$)
[dashed lines in Fig.~\ref{Figure_04}(b)]
are not the same.
Since the Fourier transformation of the transfer integral between $p_z$ orbitals,
$t(\Vec{q})$ [Eq.~\eqref{eq:t(q)}],
are isotropic along the in-plane direction, 
$t(\Vec{q})$ depends only on $|\Vec{q}|$.
If any electronic state is mainly comprised of three monolayer states
of which waves vectors are arranged in a 3-fold rotationally symmetric way around $K^\xi$,
e.g., $K^\xi+\Vec{k}$, $K^\xi+R(2\pi/3)\Vec{k}$, $K^\xi+R(4\pi/3)\Vec{k}$, 
then the $|\Vec{q}|$ associated with the interaction
between the monolayer states are identical
since we can freely choose one among the three equivalent $K^\xi$,
i.e., $K^\xi$, $R(2\pi/3)K^\xi$, $R(4\pi/3)K^\xi$,
for each state.
This is the case that happens at the mini Dirac point
(the hexagonal Brillouin zone corners)
of graphene on hBN \cite{PhysRevB.90.155406}.
On the contrary, $t(\Vec{q})$ for each interaction
of the resonant coupling in BN/G/BN
are not identical,
since there are more than three states involved.
As a result, $u_0^2$ in Eq.~\eqref{eq:V0_V1}
varies $\pm 2,3\%$ for $\theta_1$, $\theta_2$, $\theta_3$,
and $\pm11\%$ for $\theta_4$.
Nevertheless,
since the twelve wave vectors are arranged in a 12-fold rotationally symmetric way,
equally spaced triplets (i.e., $n\in \{0, 4, 8\}$, $n\in \{1, 5, 9\}$,
$n\in \{2,6,10\}$, $n\in \{3,7,11\}$) satisfy the 3-fold rotational symmetry with $K^\xi$.
As a result, the oscillation of $u_0^2$
is consistent with the rotational symmetry of $H_\mathrm{ring}^\xi$
and does not reduce the 3-fold symmetry further.
Hereafter, we will consider the structures with $\theta_1$, $\theta_2$, $\theta_3$ only,
and use $u_0=-t(K^\xi)\approx 0.152~\mathrm{eV}$,
i.e., ignore the variation of $u_0$.

Since a hBN layer lacks the inversion symmetry,
it is natural to ask whether the band structures change
if we rotate one of the hBN layers by $180^\circ$
(We label this structure BN/G/NB).
The BN/G/NB also belongs to the symmetry group $D_3$,
while the three $C_2$ axes are $60^\circ$ rotated from
those of BN/G/BN.
It is well known that
twisted double bilayer graphene
\cite{PhysRevB.99.235406,PhysRevB.99.235417,PhysRevX.9.031021,PhysRevB.102.035421}
and BN/G/BN
at general angles \cite{Leconte_2020}
show the change of electronic structures
with respect to such a change.
The $180^\circ$ rotation of hBN
corresponds to the swap of the boron and nitrogen atoms.
Thus, we can get the effective potential Eq.~\eqref{eq:V_hBN}
of the moir\'{e} superlattice from such a layer
by replacing $\psi$ by $-\psi+2\pi/3$ [Eq.~\eqref{eq:V0_V1}],
while keeping that of the other moir\'{e} superlattice unchanged.
However, we can reduce the Hamiltonian of BN/G/NB
to that of BN/G/BN [Eq.~\eqref{eq_dodeca_H_ring}]
by a similarity transformation that
multiplies $e^{2i\xi\psi}\omega^{-\xi}$ ($e^{-2i\xi\psi}\omega^{\xi}$)
to the Bloch bases $|\Vec{k}^{(n)}\rangle$
with $n\equiv 2,3 ~(\mathrm{mod}~4)$
for $\theta_1$, $\theta_3$, $\theta_4$ ($\theta_2$).
As a result, the resonant states are invariant with respect to
the replacement of BN to NB.

\begin{figure}[ht]
\centering
\includegraphics[width=0.9\linewidth]{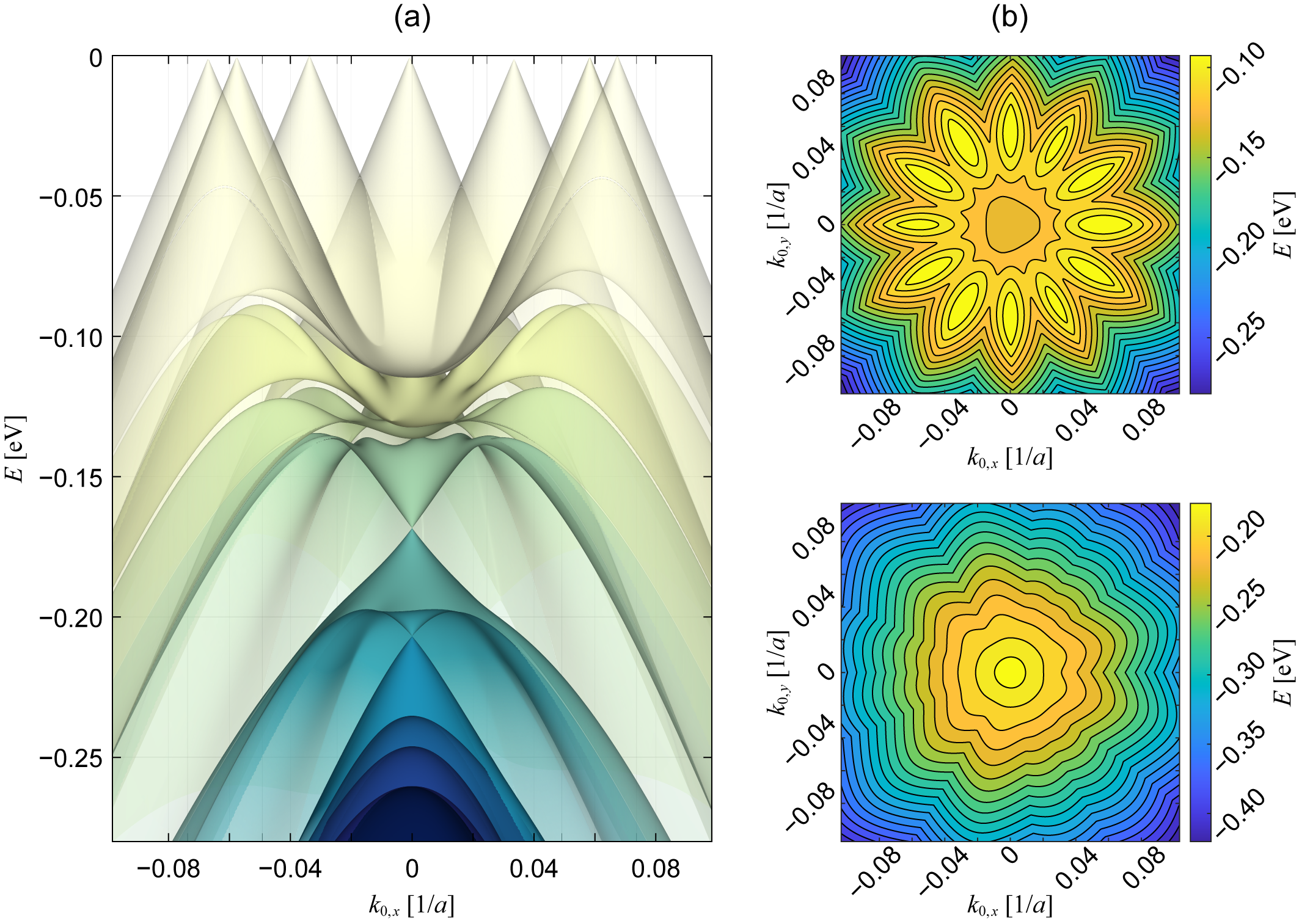}
\caption{
(a) Electronic structure in the valence band side
of the $30^\circ$ rotated double moir\'{e} superlattice at $\theta_1$.
(b) Energy contours of the
third (top panel) and the sixth (bottom panel)
valence bands
which clearly show the 3-fold rotational symmetry.
}
\label{Figure_05}
\end{figure}


Figure \ref{Figure_05}(a) shows the valence band structures
of BN/G/BN at $\theta_1$
near the resonant states at $\Vec{C}_n$
plotted as a function of $\Vec{k}_0$.
The twelve Dirac cones are arranged on a circle with a radius
$\Delta k = 2|\Vec{G}_i^{\mathrm{M},(l)}| \sin(\pi/12)$
and they are strongly hybridized near $\Vec{k}_0=\Vec{0}$
to exhibit the characteristic dispersion
including parabolic bottoms, a frilled band edge,
and new Dirac points at $-0.164~\mathrm{eV}$ and $-0.203~\mathrm{eV}$.
We have similar resonant states also in the conduction band,
while the energy spacing between the resonant states
is much smaller than in the valence band,
just like the cases of graphene on hBN \cite{PhysRevB.90.155406}
and quasicrystalline twisted bilayer graphene \cite{PhysRevB.99.165430}.
As predicted by the symmetry of $H_\mathrm{ring}^\xi$,
the band structures exhibit three-fold rotational symmetry
around $\Vec{k}_0=\Vec{0}$,
as we can clearly see from the energy contours
in Fig.~\ref{Figure_05}(b).
The structures with the other angles, $\theta_i$ ($i=2,3,4$),
also show similar band dispersion,
except that the resonant states are formed at the energies
far from the charge neutrality point of graphene,
since they have longer $|K^\xi-\Vec{C}_n|$.
The energy splitting between the resonant states in BN/G/BN, $\mathcal{O}(|V_1|)$,
is much smaller than that in the quasicrystalline twisted bilayer graphene, $\mathcal{O}(|u_0|)$,
since the interaction here involves a second order scattering
through the hBN layer.
Nevertheless, the resonant states of BN/G/BN
appear at the energies much closer to the charge neutrality point of graphene
than those of quasicrystalline twisted bilayer graphene (about $\pm1.7~\mathrm{eV})$,
since the wave vectors responsible for the interaction in BN/G/BN 
($\mathcal{O}(|\Vec{G}_i^{\mathrm{M},(l)}|)$)
are much shorter than those in quasicrystalline twisted bilayer graphene
($\mathcal{O}(|\Vec{a}_i^*|)$).
Thus, the resonant states of BN/G/BN appear
at much smaller, experimentally feasible, electron densities.

At $\Vec{k}_0=\Vec{0}$, 
we can reduce $H_\mathrm{ring}^\xi$ to an $8\times 8$ form,
\begin{align}
H_{\rm ring}^{(\xi,m)}(\Vec{k}_0) = 
\begin{pmatrix}
H_0 & W_2^\xi & 0 & Y_1^{\xi\dagger} \omega^{-m} \\
W_2^{\xi\dagger} & H_1 & X_2^{\xi\dagger} & 0\\
0 & X_2^\xi & H_2 & W_1^{\xi\dagger} \\
Y_1^\xi \omega^m  & 0& W_1^\xi & H_3 \\
\end{pmatrix},
\label{eq_8x8_H_ring}
\end{align}
by using the Bloch condition along the one-dimensional chain.
Here, $H_i=H_i^{(0)}(\Vec{k}_0=\Vec{0})$ ($i=0,1,2,3$)
and $m=-1,0,1$ is the quantized angular momentum respecting the 3-fold rotational symmetry.
The Hamiltonian $H_\mathrm{ring}^{(\xi,m)}$ exhibits a symmetry
\begin{equation}
(\Sigma \mathcal{K})^{-1} \; H_\mathrm{ring}^{(\xi,m)} ~ \Sigma \mathcal{K} = H_\mathrm{ring}^{(\xi,m')},
\end{equation}
for $m$ and $m'$ satisfying $m+m' \equiv -\xi ~(\mathrm{mod}~3)$.
Here, $\Sigma$ is $\mathrm{diag}(\sigma_x\chi,\sigma_x,\sigma_x\chi^*,\sigma_x)$
for $\theta_1$, $\theta_3$, $\theta_4$ and
$\mathrm{diag}(\sigma_x,\sigma_x\chi^*,\sigma_x,\sigma_x\chi)$
for $\theta_2$, where
$\chi$ is $e^{2i\xi\psi}\omega^{-\xi}$, and
$\mathcal{K}$ stands for complex conjugation.
Thus, the states with $(m,m')=(0,-\xi)$ form twofold doublets,
and belong to two-dimensional $E$ irreducible representation
of $D_3$ point group,
while the states $m=\xi$ is non-degenerate, 
and belong to either of $A_1$ or $A_2$.

Figure \ref{Figure_06} shows the quasicrystalline wave functions
of the third resonant state ($m=\xi$) in the hole side
at $\Vec{k}_0=\Vec{0}$ on the graphene lattice.
Figures \ref{Figure_06}(a)-(c) show the wave functions
in a system in the absence of the inversion symmetry,
i.e., $V_\mathrm{N}\ne V_\mathrm{B}$
and $\psi \neq \pi/3 ~(\mathrm{mod}~\pi)$,
such as graphene encapsulated by hBN layers,
and Figs.~\ref{Figure_06}(d)-(f) show those
in a system with the inversion symmetry,
i.e., $V_\mathrm{N} = V_\mathrm{B}$
and $\psi \equiv \pi/3 ~(\mathrm{mod}~\pi)$,
such as graphene encapsulated by
a material having the same types of atoms in both sublattices.
In both systems, the wave amplitude show the quasicrystalline order which
is distributed
on a limited number of sites
in a pattern which is incompatible with the periodicity.
Unlike the rotational symmetry of the double moir\'{e} pattern (12-fold), however,
the wave functions in the absence of the inversion symmetry
show a 3-fold rotational symmetry,
while those in the presence of the inversion symmetry
show a 6-fold rotational symmetry,
which slightly resembles the probability distribution
of the system with a true 12-fold rotational symmetry \cite{PhysRevB.99.165430}.
It should be noted that, however,
the wave functions of the system with the inversion symmetry
lose the 6-fold rotational symmetry at $\Vec{k}_0\ne \Vec{0}$.
Figures \ref{Figure_06}(a) and (d), (b) and (e), (c) and (f)
show the wave functions at $\theta_1$,$\theta_2$,$\theta_3$, respectively.
The length scale of the patterns is much larger than
that of the quasicrystalline wave functions in quasicrystalline twisted bilayer graphene \cite{PhysRevB.99.165430},
since the difference between wave vectors involved
is on the order of $|\Vec{G}_i^{\mathrm{M},(l)}|$ in the dual moir\'{e} superlattice
and on the order of $|\Vec{a}_i^*|$ in the twisted bilayer graphene
($|\Vec{G}_i^{\mathrm{M},(l)}| \ll |\Vec{a}_i^*|$).
In addition, the systems with $\theta_1$ and $\theta_2$
show larger scale,
since $|\Vec{G}_i^{\mathrm{M},(l)}|$ is almost proportional to $\theta$
[Fig.~\ref{Figure_03}(c)].
Such an electron distribution would be prominent
at the energies where the band curvature of the resonant states
are large enough to give high density of states.
At the energies away from the resonant states, on the contrary,
we will mainly see the simple overlap of the periodic wave functions of
the two typical graphene on hBN superlattice with a twist angle of $30^\circ$.




\section*{Conclusions}

We investigated the lattice configuration
and electronic structures of a double moir\'{e} superlattice
of which two hexagonal \textit{moir\'{e} patterns}
are arranged in a dodecagonal quasicrystalline configuration.
We first find the condition which gives
a $30^\circ$ stack of the two moir\'{e} patterns
in graphene encapsulated by another layers,
and show that there are 0 to 4 such configurations
depending on the lattice mismatch between
graphene and the encapsulating layer.
And we show that,
although the moir\'{e} patterns satisfy a 12-fold rotational symmetry,
the actual atomic lattice has only a 3-fold rotational symmetry ($D_3$)
if the encapsulating layers have different atomic species in the sublattices
(e.g., hBN).

We then reveal the resonant interaction
which brings together and hybridize twelve degenerate Bloch states of monolayer graphene
as well as the band dispersion around the resonant states.
Compared to the resonant states
of quasicrystalline twisted bilayer graphene,
of which \textit{atomic lattices} are arranged in a dodecagonal configuration,
the resonant states of double moir\'{e} superlattice
lack the 12-fold rotational symmetry;
they hexagonal quasicrystalline order at a specific point $\Vec{k}_0=\Vec{0}$
in the Brillouin zone
if the encapsulating layers the same types of atoms in both sublattices,
and trigonal quasicrystalline order otherwise.
These unique states appear at the energies
much closer to the charge neutrality point of graphene
and experimentally feasible electron densities.

\bibliography{twistronics}

\begin{thebibliography}{10}
\bibitem{comment:tau}
\bibinfo{note}{Note that we defined the sublattice coordinates
  $\boldsymbol{\tau}_X$ ($X=A,B$) and $\boldsymbol{\tau}_{\tilde{X}^{(l)}}$
  ($\tilde{X}=N,B$, $l=t,b$) differently from those in our previous work
  \cite{PhysRevB.90.155406}, to make the points with the highest rotational
  symmetry, i.e., the hexagonal center, as the center of the system. However,
  both definitions keep the interaction matrices [Eq.~\eqref{eq:W0-W3}] the
  same.}
\bibitem{comment:for_atomic_layers_other_than_hBN}
\bibinfo{note}{Note that the theoretical model can be easily expanded to the
  systems with the encapsulating layers other than hBN; e.g., it is
  straightforward to expand the model to the atomic layers of which energy
  bands are close to the charge neutrality point of graphene by explicitly
  using the Bloch bases for those layers.}
\end{thebibliography}

\section*{Acknowledgements}

J.A.C was supported by the National Science Foundation of China (Grant No.~12050410228).
P.M. acknowledges the support by
National Science Foundation of China (Grant No.~12074260) and
Science and Technology Commission of
Shanghai Municipality (Shanghai Natural Science Grants,
Grant No.~19ZR1436400).
J.A.C. and P.M. were supported by
the NYU-ECNU Institute of Physics at NYU Shanghai. This
research was carried out on the High Performance Computing
resources at NYU Shanghai.

\section*{Author contributions statement}
J.A.C. and P.M. conducted the calculation, analysis, and wrote the manuscript.

\section*{Additional information}

\textbf{Competing interests}
The authors declare no competing interests.

\end{document}